\documentclass[conference]{IEEEtran}

\usepackage{graphicx}    
\usepackage{float}       
\usepackage{url}         
\usepackage{caption}     
\usepackage{subcaption}  
\usepackage{cite}        
\usepackage{marvosym}
\graphicspath{{./figures/}}
\DeclareGraphicsExtensions{.pdf,.jpg,.png}

\title{Poster: SpiderSim: Multi-Agent Driven Theoretical Cybersecurity Simulation for Industrial Digitalization}

\author{
    \IEEEauthorblockN{
        Jiaqi Li\IEEEauthorrefmark{1},
        Xizhong Guo\IEEEauthorrefmark{2},
        Yang Zhao\IEEEauthorrefmark{3},
        Lvyang Zhang\IEEEauthorrefmark{4},
        Lidong Zhai\IEEEauthorrefmark{5}\textsuperscript{\Letter}
    }
    \IEEEauthorblockA{
        \IEEEauthorrefmark{1}\IEEEauthorrefmark{2}\IEEEauthorrefmark{3}\IEEEauthorrefmark{4}\IEEEauthorrefmark{5}Institute of Information Engineering, Chinese Academy of Sciences, Beijing, China \\
        \IEEEauthorrefmark{1}\IEEEauthorrefmark{2}\IEEEauthorrefmark{3}\IEEEauthorrefmark{4}School of Cyber Security, University of Chinese Academy of Sciences, Beijing, China \\
    }
}

\begin{document}

\maketitle

\begin{abstract}
Rapid industrial digitalization has created intricate cybersecurity demands that necessitate effective validation methods. While cyber ranges and simulation platforms are widely deployed, they frequently face limitations in scenario diversity and creation efficiency. In this paper, we present SpiderSim, a theoretical cybersecurity simulation platform enabling rapid and lightweight scenario generation for industrial digitalization security research. At its core, our platform introduces three key innovations: a structured framework for unified scenario modeling, a multi-agent collaboration mechanism for automated generation, and modular atomic security capabilities for flexible scenario composition. Extensive implementation trials across multiple industrial digitalization contexts, including marine ranch monitoring systems, validate our platform's capacity for broad scenario coverage with efficient generation processes. Built on solid theoretical foundations and released as open-source software, SpiderSim facilitates broader research and development in automated security testing for industrial digitalization.

keywords: Cybersecurity simulation; atomic capabilities; industrial digitalization; multi-agent; theoretical simulation
\end{abstract}



%

\section{Introduction}
The rapid advancement of industrial digitalization has introduced unprecedented cybersecurity challenges across manufacturing, energy, transportation, and other sectors. Attack-defense simulations have proven effective for training and validating security capabilities. Frameworks like MITRE ATT\&CK\cite{ATTCK} and Engage\cite{engage} facilitate defense improvements. 
While traditional cyber ranges and security testbeds provide valuable validation environments, they often face limitations in scenario coverage and generation efficiency. First, traditional cyber ranges require substantial resources for environment setup and maintenance. Second, existing platforms often lack systematic methodologies for rapid scenario generation. Third, the manual scenario creation process limits comprehensive coverage of security situations in industrial digitalization contexts. As an open-source network simulation platform, CyberBattleSim\cite{cbs} utilizes abstract modeling and reinforcement learning to advance the field. However, it focuses narrowly on internal networks rather than capturing real-world complexities. The platform also implements limited techniques and lacks integration with security tools.
To address these challenges, we present SpiderSim, a theoretical cybersecurity simulation platform that introduces three key innovations. First, it implements a structured framework for unified scenario modeling, which enables standardized scenario construction through formal methodologies. Second, it incorporates a multi-agent collaboration mechanism for automated scenario generation and validation. Third, it provides modular atomic security capabilities that support flexible composition of industrial security scenarios. This integrated approach achieves both scenario diversity and generation efficiency without compromising theoretical precision.

The remainder of this paper is organized as follows: Section II presents the platform architecture and core components of SpiderSim. Section III details the implementation and application through industrial digitalization cases. Section IV concludes the paper and discusses future research directions.

\begin{figure}[h]
    \centering
    \includegraphics[scale=0.25]{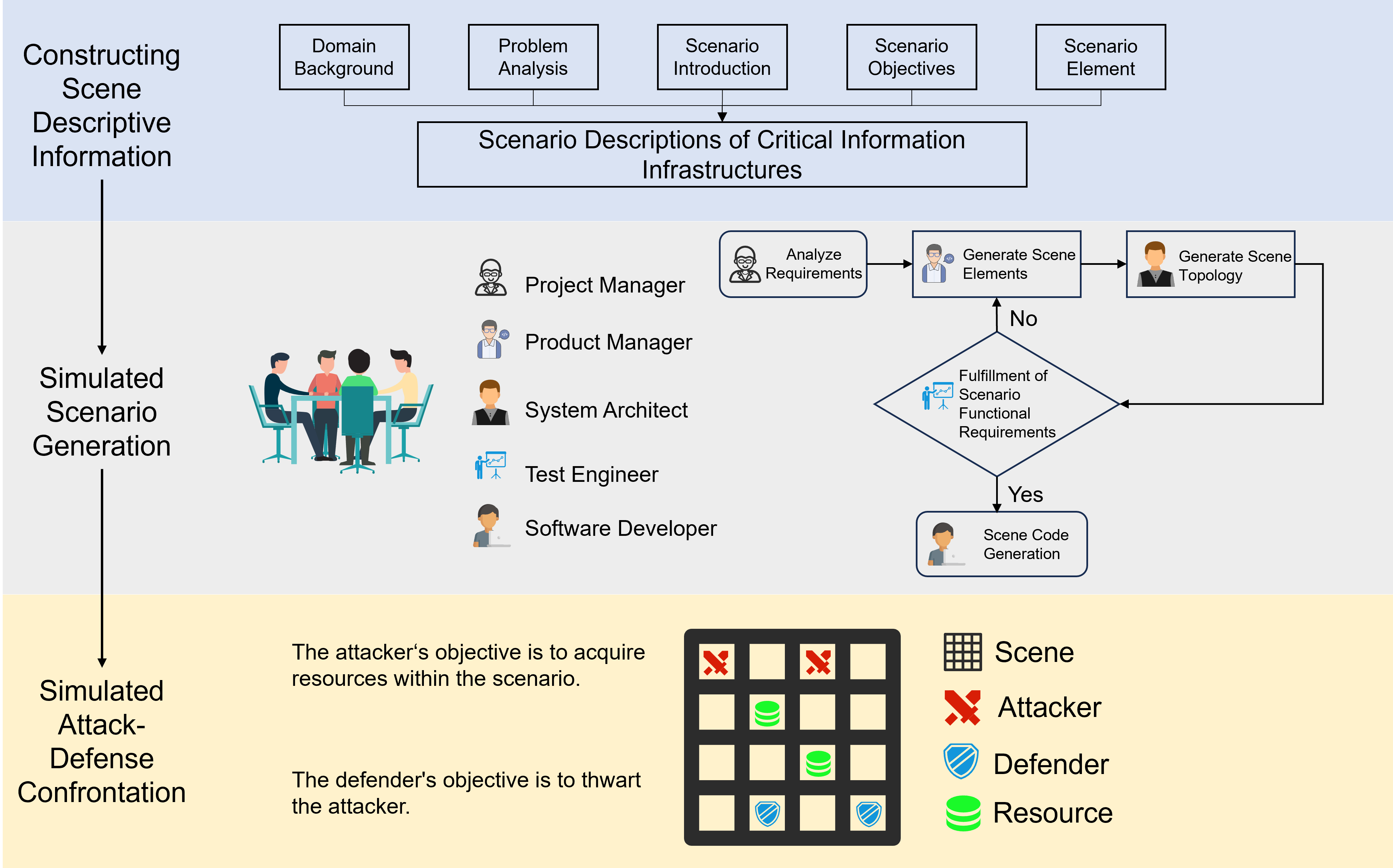}
    \caption{Automated scenario generation and experimental framework}
    \label{fig:tuopu}
\end{figure}

\section{PLATFORM ARCHITECTURE AND COMPONENTS }
SpiderSim implements a three-layered architecture that systematically transforms abstract security requirements into executable attack-defense scenarios. The platform integrates a unified scenario modeling framework for structured requirement specification, a multi-agent collaboration mechanism for automated scenario development, and atomic security capabilities for comprehensive security validation. This hierarchical design enables efficient scenario generation while preserving theoretical rigor and practical applicability.
The layered architecture directly addresses the fundamental limitations of existing cybersecurity simulation platforms, including complex environment configuration requirements, time-consuming manual scenario creation processes, and insufficient coverage of industrial security contexts. By emphasizing lightweight deployment, automated generation workflows, and modular security components, SpiderSim provides a flexible and efficient solution for industrial security research and validation through three core components.

\subsection{Unified scenario modeling framework}
The foundational component of SpiderSim rests upon a systematic methodology for scenario construction. At its core, the framework encompasses comprehensive domain context analysis, structured problem decomposition, detailed scenario specifications, clear objective definitions, and essential element composition. Built on a formalized approach, it ensures consistent scenario quality while enabling rapid generation across diverse industrial digitalization contexts. Such methodological structure supports both standardization and customization, allowing for efficient scenario creation while maintaining contextual relevance.

\subsection{Multi-agent collaboration mechanism}
An automated generation engine drives scenario development through coordinated agent interactions.  Agents operate through synchronized communication channels, coordinating scenario development activities from initial requirement analysis through final validation. This mechanism incorporates continuous refinement processes, enabling dynamic adjustments based on validation results and emerging requirements. The collaborative approach ensures scenario completeness while maintaining generation efficiency through automated coordination and validation procedures.

\subsection{Atomic security capabilities}
SpiderSim supports the construction of complex network attack and defense capabilities using atomized modules. These modules include the Shocktrap module\cite{hong}, honeypot module, vulnerability scanning module, among others. Users can select the necessary modules to build a customized security defense system or test the effectiveness of individual modules. Meanwhile, SpiderSim incorporates Cybersecurity Chess Manual \cite{wuposter}, to help generate more atomized attack and defense scenarios.
For example, users can choose a data encryption module, honeypot module, and Shocktrap module to form a security strategy to counteract phishing attacks. This atomized and reconfigurable module design enhances the flexibility of SpiderSim in constructing diverse attack and defense scenarios. Standardized interfaces between modules also allow third parties to develop and submit new attack and defense modules to expand the platform's capabilities.

\section{PRACTICAL IMPLEMENTATION AND CASE STUDIE}
Utilizing the aforementioned framework, we constructed a digital environment for a marine ranch.  Figure 2(a) illustrates the framework for the marine ranch monitoring system, which encompasses devices such as sensors, control networks, video surveillance systems, and remote maintenance systems. Figure 2(b) displays the theoretical-level network topology and attack path diagram constructed within SpiderSim. Within this environment, we conducted a series of cyber attack-defense experiments, which led to the development of a security protection scheme tailored to the typical threats faced by the system. This scheme has been tested and proven to effectively mitigate the risk of cyber attacks on the marine ranch infrastructure.

\begin{figure}[h]
    \begin{subfigure}[b]{0.23\textwidth}
        \centering
        \includegraphics[scale=0.2]{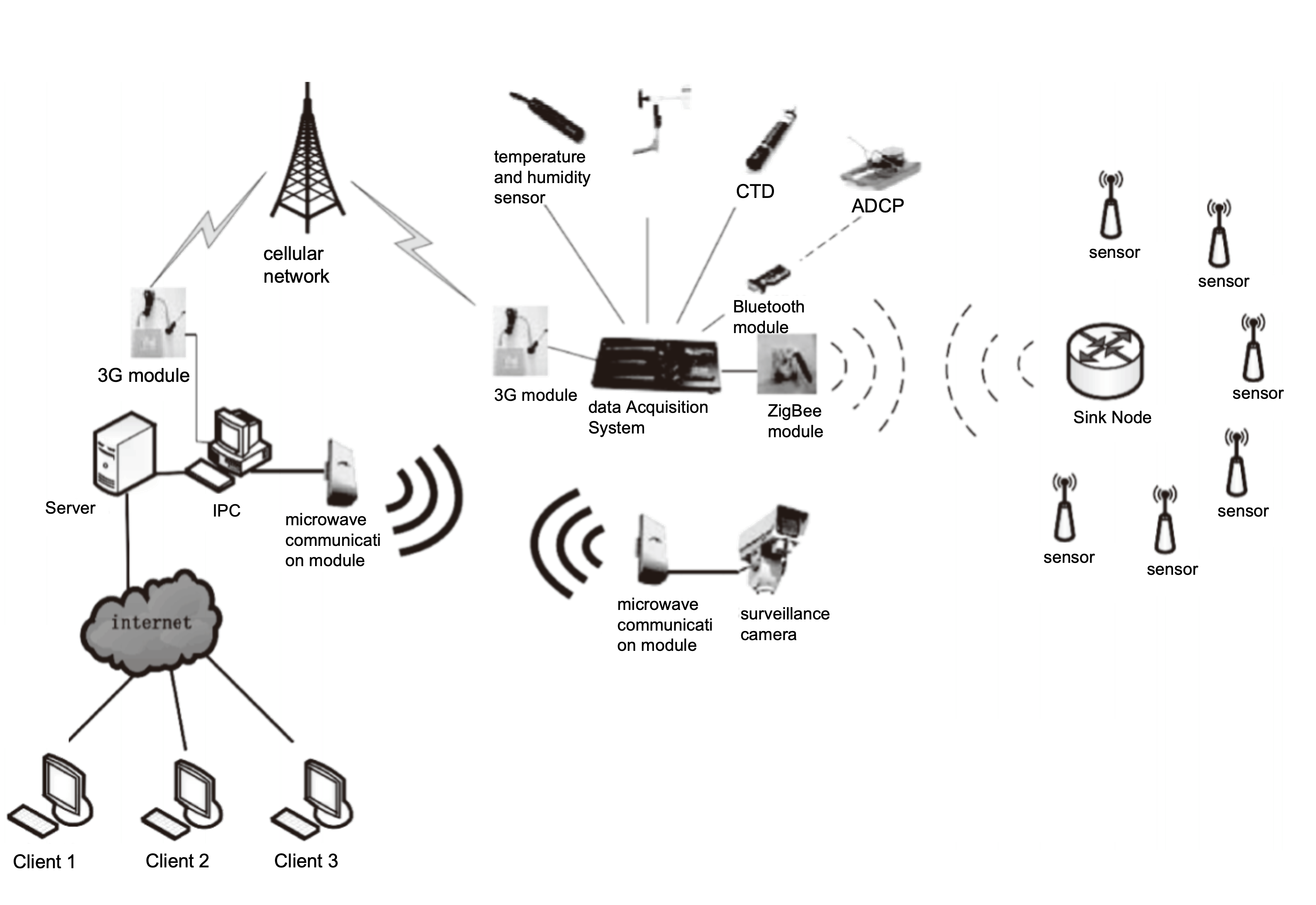}
        \caption{The framework for the marine ranch monitoring system}
        \label{fig:tuopu}
    \end{subfigure}
    \hspace{0.5cm}
    \begin{subfigure}[b]{0.23\textwidth}
        \centering
        \includegraphics[scale=0.3]{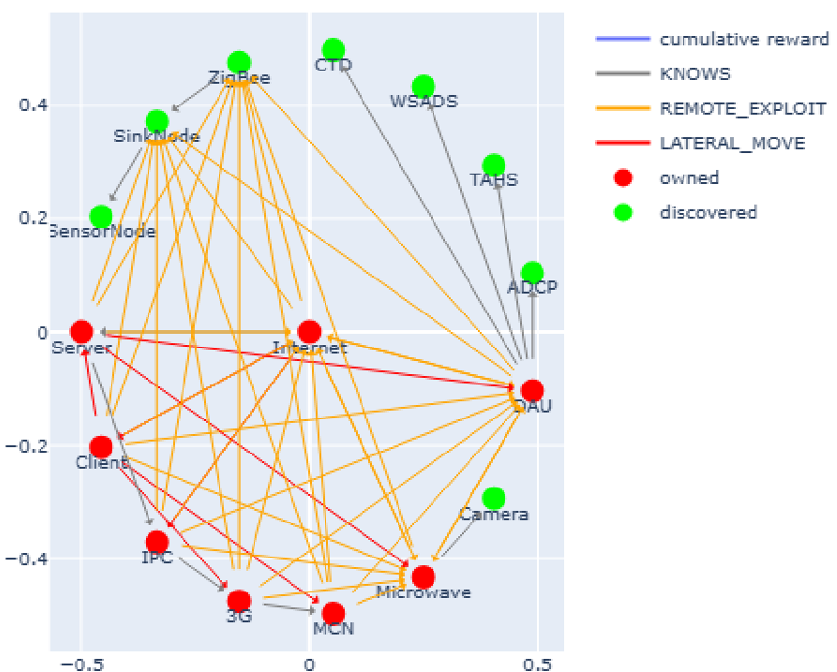}
        \caption{The network topology and attack path diagram}
        \label{fig:frameword}
    \end{subfigure}
    \caption{Digital environment and cyber attack-defense experiments of marine ranch}
    \label{fig:overall}
\end{figure}

\section{Conclusion}
SpiderSim represents an innovative approach to theoretical security simulation in industrial digitalization contexts. The platform successfully implements rapid scenario generation through systematic scene description and multi-agent collaboration, as demonstrated through practical testing. The lightweight architecture demonstrates particular advantages in complex industrial contexts where rapid scenario development is critical. 

The open-source implementation (available at \url{https://github.com/NRT2024/SpiderSim}) provides a foundation for collaborative development of next-generation security testing solutions for industrial digitalization.







%



\bibliographystyle{plain}
\bibliography{mypaper}

\newpage
\thispagestyle{empty}
\begin{figure}[h]
  \centering
  \includegraphics[width=\textwidth]{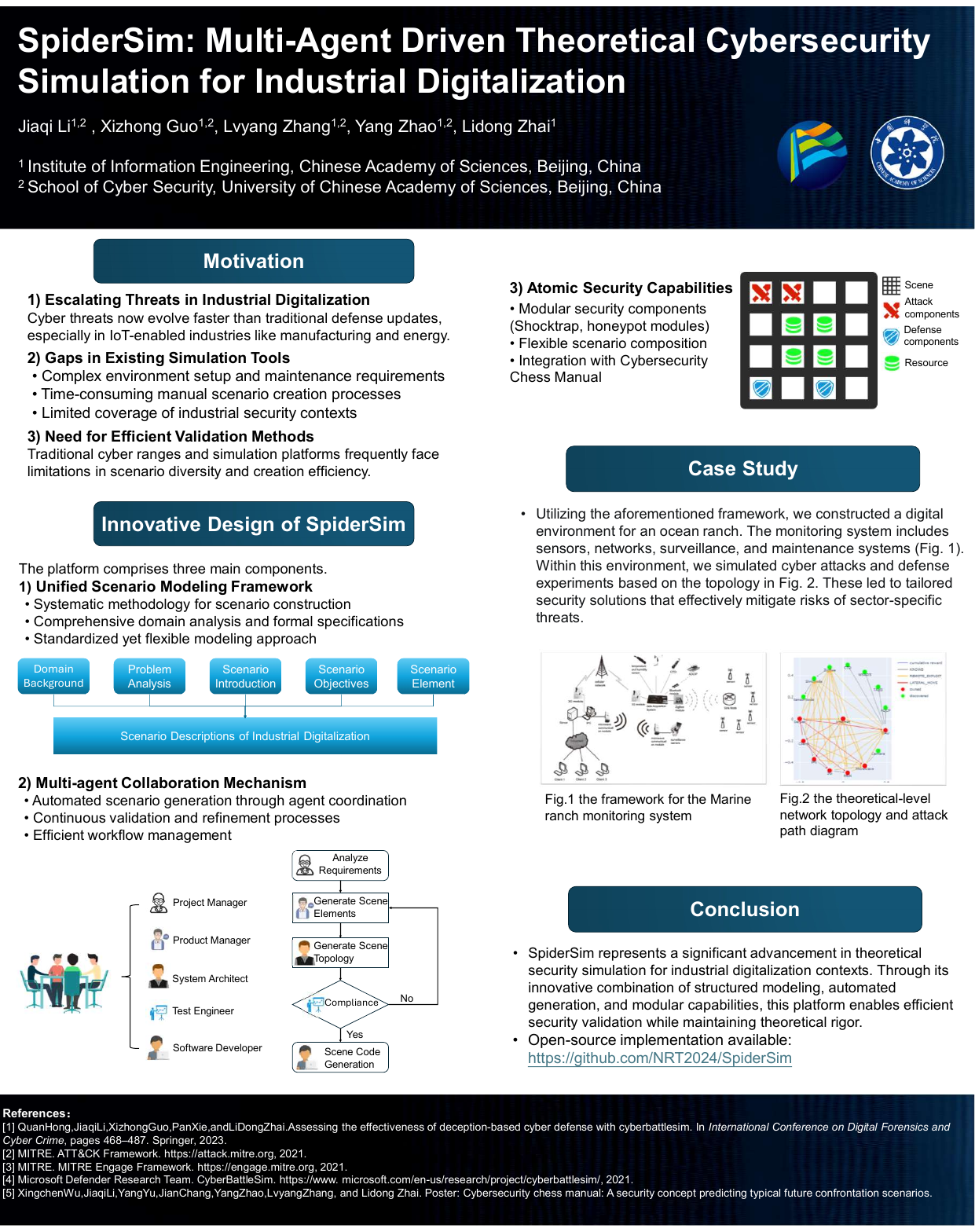}
\end{figure}

\end{document}